\documentclass{elsart5p}
\usepackage{amssymb}
\usepackage{graphicx}  
\usepackage{wasysym}
\usepackage{lineno}

\begin{document}
\begin{frontmatter}

\title{Precision efficiency calibration of a high-purity co-axial germanium detector at low energies}
\author[label1]{B. Blank$^1$,}
\author[label1]{P. Ascher,} 
\author[label1]{M. Gerbaux,} 
\author[label1]{J. Giovinazzo,} 
\author[label1]{S. Gr\'evy,} 
\author[label1]{T. Kurtukian Nieto,} 
\author[label1]{M. Versteegen,} 
\author[label2]{J.C. Thomas}

\address[label1]{Centre d'Etudes Nucl\'eaires de Bordeaux Gradignan - 
        UMR 5797, CNRS/IN2P3 - Universit\'e de Bordeaux, 19, Chemin du Solarium, CS 10120,
        F-33175 Gradignan Cedex, France}
\address[label2]{Grand Acc\'el\'erateur National d'Ions Lourds, 
        CEA/DRF - CNRS/IN2P3, Bvd Henri Becquerel, BP 55027, F-14076 CAEN Cedex 5, France}
        
\begin{abstract}

Following work done in the energy region above 100 keV, the high-precision calibration of a co-axial high-purity 
germanium detector has been continued in the energy region
below 100~keV. Previous measurements or Monte-Carlo simulations have been repeated with higher statistics and new source
measurements have been added. A precision as in the high-energy part, i.e. an absolute precision for the detection 
efficiency of 0.2\%, has been reached. The low-energy behaviour of the germanium detector was further scrutinized by
studying the germanium X-ray escape probability for the detection of low-energy photons.
In addition, one experimental point, a $\gamma$\ ray at 2168~keV from the decay 
of $^{38}$K, has been included for the total-to-peak ratios agreeing well with simulations. The same $\gamma$ ray
was also added for the single- and double-escape probabilities. Finally, the long term stability of the efficiency of the
germanium detector was investigated by regularly measuring the full-energy peak efficiency with a precisely calibrated $^{60}$Co
source and found to be perfectly stable over a period of 10 years.

\end{abstract}

\end{frontmatter}


\footnotetext[1]{Corresponding author: B. Blank, blank@cenbg.in2p3.fr}

\setcounter{footnote}{0}

\section{Introduction}

Gamma-ray spectroscopy is a powerful tool for the study of nuclear structure. In particular, $\beta$-decay studies are
often performed by measuring $\gamma$-ray intensities to deduce $\beta$-decay feeding probabilities.
This applies in particular to high-precision $\beta$-decay measurements as conducted for the purpose of weak-interaction 
studies. To determine the $ft$ value of super-allowed $\beta$ decays of the 0$^+ \rightarrow $ 0$^+$ type, high-precision
measurements are required for the $Q_{EC}$ value, the half-life, and the super-allowed branching ratio. The latter quantity is 
usually determined by measurements of the $\gamma$-decay probabilities of excited levels in the $\beta$-decay daughter nucleus,
from which the $\beta$-decay feeding probabilities are determined.

Our group has a long standing experimental program of measuring half-life values and branching ratios of these super-allowed $\beta$ 
decays~\cite{blank04ga62,canchel05,bey08,matea08,kurtukian09,blank10ca3839,souin11,blank15ca38} or of mirror 
decays~\cite{blank10ca3839,bacquias12,magron17}, where high-precision $\gamma$-ray spectroscopy is required. Therefore, we have
started in 2010 to precisely calibrate a high-purity (HP), n-type, co-axial germanium detector with a relative efficiency of 70\%.
The most important part of this calibration work was published a few years ago~\cite{blank14ge}. In recent years, we continued this
calibration work, in particular to improve the precision below 100~keV $\gamma$-ray energy, where we gave an absolute precision
of only 1\% in our previous work. This was mainly due to the fact that at low energies, many background $\gamma$ rays "pollute" the 
spectrum and we had not enough information to ascertain that we achieved the same precision as above 100~keV.

For a measurement recently performed with $^{22}$Mg, we need precise efficiencies in the energy range below 100~keV, which was the prime reason 
to extend the high-precision efficiency calibration of our detector down to 40~keV.
In the present work, we describe new measurements with a new source ($^{169}$Yb) and with higher statistics ($^{24}$Na, $^{48}$Cr 
and $^ {207}$Bi). Higher-statistics Monte-Carlo (MC) simulations have also been performed for  $^{133}$Ba, $^{152}$Eu and 
$^{180}$Hf$^m$. 
The new experimental data for $^{24}$Na, $^{48}$Cr, and $^{169}$Yb were taken at ISOLDE.

\begin{table}[hht]
\renewcommand{\arraystretch}{1.25}
\caption{$^{169}$Yb and $^{180}$Hf$^m$ source characteristics used to determine the detector efficiency.
         The characteristics of the other sources are given in~\cite{blank14ge}.}                                                                        
\begin{center}
\begin{tabular}{ccccc}
\multicolumn{5}{c}{} \\
\hline \rule{0pt}{1.3em}
~~Nuclide~~& T$_{1/2}$~~   &~~ $E_\gamma$ (keV) ~~  & ~~$P_\gamma$ (\%) & ~~Reference~~  \\
[0.5em] \hline \rule{0pt}{1.3em}
$^{169}$Yb &  32.016 d     & ~50.4                  & 1.464(21)~~       & \cite{iaea07}  \\ 
           &               & ~58.0                  & 0.3855(81)~       &                \\ 
           &               & ~63.1                  & 0.4405(24)~       &                \\                                                    
           &               & ~93.6                  & 0.02571(17)       &                \\                                                    
           &               & 109.8                  & 0.1736(9)~~       &                \\                                                  
           &               & 118.2                  & 0.01870(10)       &                \\                                                  
           &               & 130.5                  & 0.1138(5)~~       &                \\                                                   
           &               & 177.2                  & 0.2232(10)~       &                \\                                                  
           &               & 198.0                  & 0.3593(12)~       &                \\                                                  
           &               & 307.7                  & 0.10046(45)       &                \\                                                  
$^{180}$Hf$^m$ &  5.53 h   & ~57.5                  &  0.4799(94)       & \cite{ensdf}\\
           &               & ~93.3                  &  0.1751(14)       & \cite{helmer03}\\
           &               & 215.3                  &  0.8150(15)       & \cite{helmer03}\\
           &               & 332.3                  &  0.9443(05)       & \cite{helmer03}\\
           &               & 443.1                  &  0.8180(130)      & \cite{helmer03}\\
           &               & 500.7                  &  0.1421(28)       & \cite{ensdf}\\
[0.5em] \hline            
\end{tabular}             
\label{tab:sources}         
\end{center}                
\end{table}

We took also data at ISOLDE for $^{49}$Cr, which, according to literature, should have had well determined branching ratios for $\gamma$ 
rays between 62~keV and 153~keV, ideal to connect the region above 100~keV already well calibrated in our previous work and the region
below 100~keV. However, it turned out that the branching ratios were not at all in agreement with the $\gamma$-ray rates we observed.
Therefore, instead of using $^{49}$Cr to calibrate our detector, we used this latter to propose a new measurement of the branching ratios
of this isotope~\cite{blank18cr49}.

The general outcome of the present work is that the detector model developed in our previous work is also valid with the same 
precision below 100~keV. Only for data points below 40~keV we seem to start having systematic discrepancies between experimental 
data and simulations. It is not clear whether this is a problem with our detector model or rather linked to electronics issues
with the trigger efficiency. We refrain from speculating about the reason for this discrepancy and assume that the efficiency 
of our detector is precise at a level of 0.2\% down to a $\gamma$-ray energy of 40~keV. The detector model used for the simulations
in the present work is the same as in \cite{blank14ge}.

\section{Experimental results with $\gamma$-ray sources and comparison with CYLTRAN simulations}

Figure~\ref{fig:eff} shows the global results of our efficiency calibration work. In the present work, 
we have adopted the branching ratios for $^{180}$Hf$^m$ proposed by 
Helmer {\it et al.}~\cite{helmer03} and from ENSDF~\cite{ensdf} in our analysis and in the decay scheme simulations, 
because we found them in much better agreement with our data than those used previously~\cite{blank14ge}. 
The branching ratios used in the present work for $^{169}$Yb and $^{180}$Hf$^m$ are given in Table~\ref{tab:sources}. 
Part a) of figure~\ref{fig:eff} gives the absolute efficiency as a 
function of energy established with 15 different radioactive sources, standard sources commercially available, but also short-lived 
sources we produced at ISOLDE or IPN Orsay. In part b), we give the residuals between the experimental data and the MC simulations 
with the CYLTRAN code~\cite{cyltran}. The vast majority of data lies within $\pm$2\%. The residuals are consistent with no difference
between the measurements and the simulations, which is expected as the absolute normalisation is a free parameter for all sources 
except for the $^{60}$Co source the activity of which has a precision of about 0.1\%. To verify that the detection efficiency
is correctly described by our detector model, one has to inspect the residuals for a given source and check that there
is a zero difference between experimental and simulation data for low- and high-energy data points.

\subsection{Detection efficiency below 100 keV}

\begin{figure*}[hht]
\begin{center}
\includegraphics[width=0.62\textwidth,angle=-90]{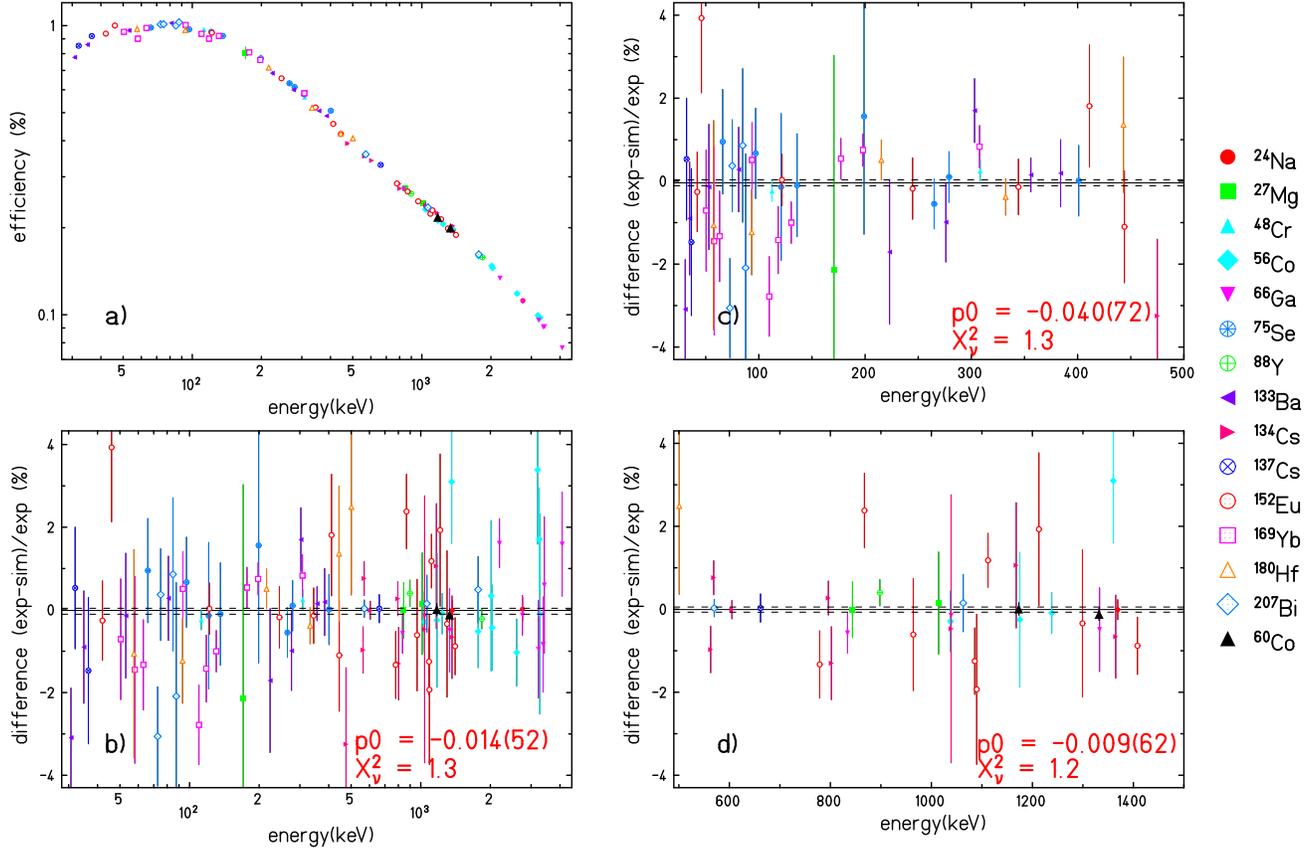}
\caption{Full-energy efficiencies and residuals for the $\gamma$-ray detection efficiency for all multi-$\gamma$-ray sources measured with the present detector.
         Compared to our previous publication~\cite{blank14ge}, higher-statistics measurements have been performed for the $^{24}$Na, 
         $^{48}$Cr and $^ {207}$Bi sources. As an additional source, $^{169}$Yb has been added for its low-energy $\gamma$ rays.
         Part a) shows the absolute efficiency of the germanium detector over the full range of 30~keV to 4~MeV.
         Part b) gives the residuals with respect to the MC calculations with the CYLTRAN code over the same energy range.
         In part c), we zoom on the energy range from 30~keV to 500~keV, whereas part d) shows the residuals in the region from 500~keV to
         1500~keV. The fit for this latter figure was made over the range 500~keV - 4000~keV. See text for details.
         } 
\label{fig:eff}
\end{center}
\end{figure*}

A fine analysis of the residuals can give more insights about the
precision of our detector model in different regions of energy. As a first step, we present in part c) and d) of figure~\ref{fig:eff} 
the residuals for $\gamma$-ray energies below and above 500~keV, respectively. The data are the same as in part b) of the figures, 
in particular no change in normalisation, whereas the fit is performed only on the respective parts of the data.
As indicated in the figures, the residuals are in agreement 
with no difference between experiment and simulations. For the three residuals plots, the $\chi^2_\nu$ is close to unity.

\begin{table}[hht]
\renewcommand{\arraystretch}{1.25}
\caption{Residuals between the experimental efficiency and the simulated one. For the present table we analysed different regions of energies.
         }                                                                        
\begin{center}
\begin{tabular}{cc}
\multicolumn{2}{c}{} \\
\hline \rule{0pt}{1.3em}
~~Energy range~~& residuals (\%)    \\
[0.5em] \hline \rule{0pt}{1.3em}
  30 - 500 keV  & -0.058(072)  \\
  30 - 400 keV  & -0.030(100)  \\
  30 - 300 keV  & -0.184(125)  \\
  30 - 200 keV  & -0.215(141)  \\
  30 - 100 keV  & -0.397(288)  \\
  30 - ~60 keV  & -0.593(463)  \\
  40 - 500 keV  &  0.002(099)  \\
  40 - 400 keV  & -0.001(100)  \\
  40 - 300 keV  & -0.143(127)  \\
  40 - 200 keV  & -0.166(144)  \\
  40 - 100 keV  & -0.010(351)  \\
  40 - ~60 keV  & -0.032(614)  \\
[0.5em] \hline            
\end{tabular}             
\label{tab:residuals}         
\end{center}                
\end{table}

A more refined analysis is presented in table~\ref{tab:residuals} for the low-energy part of our measurements. We selected 
different ranges of energies and determined whether or not the residuals are still in agreement with zero or not. Due to the energy selection, 
this analysis includes only part of the $\gamma$ rays for the different sources. No change of the normalisation of source activities was made. 
If a deviation from zero is found, this means that for one or several sources less agreement is found e.g. for low energies. 

This is the case for energies below 40~keV. Although the error bars increase significantly when using less and less data, 
there seems to be a general tendency of more and more negative residuals for the low-energy part. This indicates that there is 
missing experimental decay strength, which could come from a problem with the trigger probability for these low
$\gamma$-ray energies. Another possible explanation could be that the detector model we use is wrong with e.g.
the detector entrance window being too thin. However, in order to match experiment and simulation, we would need to increase the 
window thickness by more than 50\%, which is in contradiction with other measurements e.g. with an electron beam~\cite{blank14ge}.

Therefore, we decided to limit our conclusions to energies larger than 40~keV. Above this energy we have very good agreement between
model calculations and experiment. As we do not find any difference in agreement between the low-energy part (40 - 100~keV) and
the high-energy part treated in reference~\cite{blank14ge}, we conclude that the absolute precision on the efficiency of the 
germanium detector is 0.2\% over the energy range of 40~keV to 4~MeV, with a relative efficiency precision of 0.15\% over this
energy range.

\begin{figure}[hht]
\begin{center}
\includegraphics[width=0.35\textwidth,angle=-90]{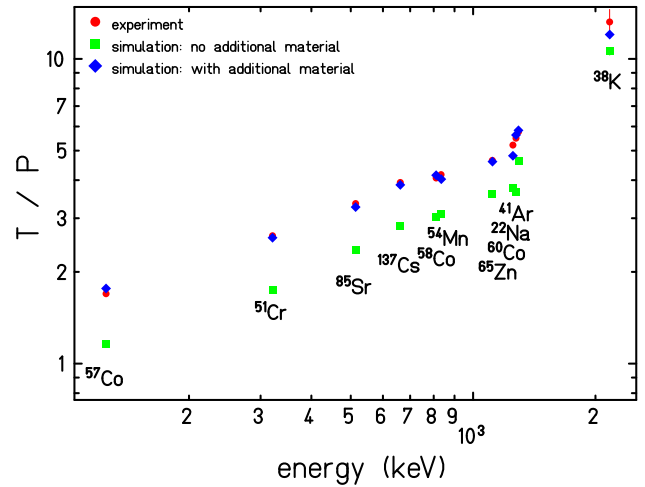}
\caption{Total-to-Peak ratio for $\gamma$-ray energies ranging from 122~keV to 2170~keV. The good agreement between experiment
         and simulation already obtained up to 1300~keV in our previous work is extended to 2170~keV.
         } 
\label{fig:t-to-p}
\end{center}
\end{figure}

\subsection{Total-to-peak ratio as well as single- and double-\-escape ratios}

As shown in figure~\ref{fig:t-to-p}, we have also updated the total-to-peak ratio plot for our detector by including a data point for $^{38}$K taken 
at ISOLDE. This additional data point is of interest, as it is much higher in energy than the other data used up to now. The new experimental data point is 
in good agreement with our detector model and confirms that matter around the detector has to be added to correctly describe the total-to-peak
ratio in our MC simulations.

\begin{figure}[hht]
\begin{center}
\includegraphics[width=0.35\textwidth,angle=-90]{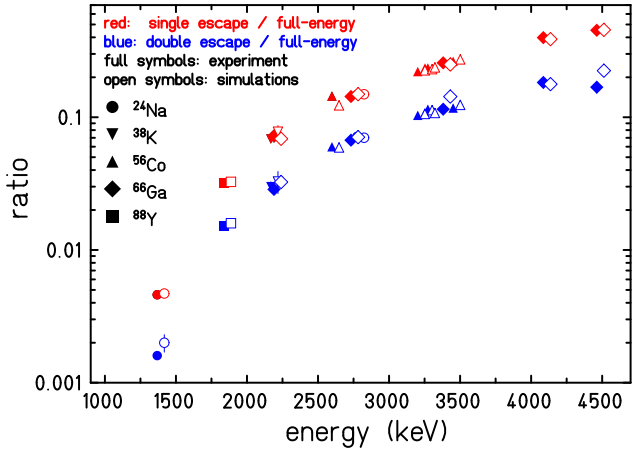}
\caption{Ratios of the single-escape (red) and  double-escape (blue) intensities over the full-energy intensities
         from source measurements with $^{24}$Na, $^{38}$K, $^{56}$Co, $^{66}$Ga, and $^{88}$Y. Full symbols represent experimental ratios, 
         open symbols, slightly shifted to facilitate the reading of the figure, are from simulations with the CYLTRAN code. The additional point for $^{38}$K fits 
         nicely in the systematics.
         } 
\label{fig:escape}
\end{center}
\end{figure}

In a similar way, we also included the $^{38}$K data in the plot of the single-escape and double-escape probabilities for the 511~keV
annihilation radiation (figure~\ref{fig:escape}). Good agreement is achieved between experiment and MC simulations.

\section{X-ray escape probability for low-energy photon detection}

An effect, which becomes important for photon energies below 100~keV is the possibility of germanium X-ray escape.
At low energies, the interaction of a photon takes place at the very entrance of the germanium crystal. When the photon interacts
by photo-electric effect or by Compton scattering, vacancies are created in the electronic shells of the germanium atoms, which
are filled by electrons from higher-lying shells thus creating X-rays. These X-rays are of low energy (of order 10~keV) and are usually 
absorbed by the germanium crystal. However, the closer
the photon interaction occurs to the surface of the germanium crystal (i.e. the lower the photon energy), the higher is the probability
of the X-rays to escape from the detector thus creating an event with less energy. Therefore, for low-energy X-ray or $\gamma$-ray 
peaks, a small satellite peak about 10~keV below the main peak can be observed, the intensity of which increases with decreasing
energy.

Helmer and co-workers~\cite{helmer03} found that their MC simulations
underestimate this effect by a factor of 1.16. After scaling the escape probability by this factor, they found satisfying 
agreement between simulations and experimental measurements. The authors argued that the assumption of a uniform detector surface
as well as of complete charge collection in the surface area might be questionable.

\begin{figure}[hht]
\begin{center}
\includegraphics[width=0.35\textwidth,angle=-90]{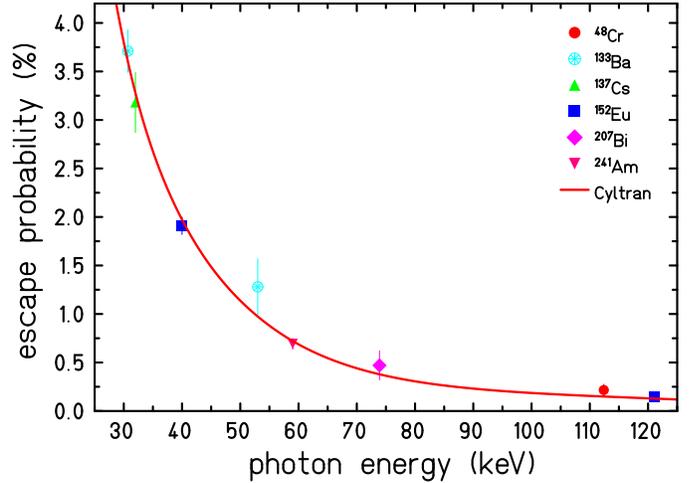}
\caption{X-ray escape probability for low-energy photons as measured with the germanium detector and simulated with the CYLTRAN code.
         The symbols give the experimental data points, whereas the full line stems from CYLTRAN simulations.} 
\label{fig:xray}
\end{center}
\end{figure}

As we demonstrate in figure~\ref{fig:xray}, we find agreement over the full energy range from 30~keV to 120~keV between the simulated
escape probability and the experimentally determined one. No scaling factor is needed. We conclude therefore that, for our detector,
CYLTRAN treats correctly this X-ray escape probability.

\section{Long-term stability of the absolute efficiency with $^{60}$Co sources}

Our germanium detector is kept cold all year long since more than 10 years (except e.g. for travel to experimental sites in Jyv\"askyl\"a or at ISOLDE). 
The reason for that is to avoid that the lithium used 
for the $p$ contacts diffuses in or out the detector and thus changes the efficiency of the detector. 

In order to verify
the efficiency of our detector, we performed several measurements over the years with a $^{60}$Co source, which is calibrated
in activity with a precision of better than 0.1\%. Data were taken with this source a last time end of 2019. As $^{60}$Co has a 
half-life of 5.271~y, the activity has now decreased too much to still use this original source. 

\begin{figure}[hht]
\begin{center}
\includegraphics[width=0.35\textwidth,angle=-90]{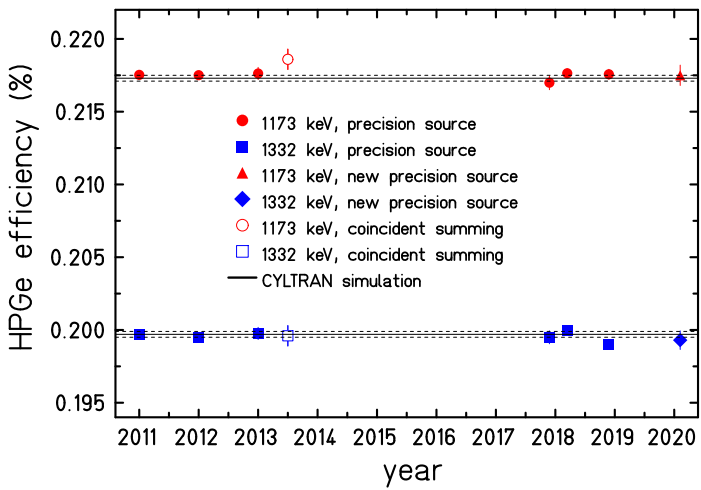}
\caption{Long-term stability of the absolute detection efficiency of the germanium detector. The efficiency is stable over a period
         of 10 years. The last data point has been measured with a new $^{60}$Co source calibrated with respect to our old source
         in August-October 2018 by means of the germanium detector presently described.
         This source, with an activity of 19.948(28)~kBq on August 2, 2018, has a precision on its activity ($\Delta$A/A~= 0.14\%) slightly 
         worse than our previous source ($\Delta$A/A~= 0.09\%). The precision of the last data point in the plot is limited by 
         the detection statistics. The full lines represent the calulated efficiencies with the dashed lines being the error bars.
         } 
\label{fig:long}
\end{center}
\end{figure}

We have therefore calibrated a new $^{60}$Co source with a precision of 0.14\%, which will be used in the future. This source has
the same mounting as the old source. The calibration was performed with respect to the old precision
source in several runs during 2018. A first verification of the detector efficiency was performed with the new source beginning of 2020
yielding agreement with the initial efficiency (see figure~\ref{fig:long}).

\section{Conclusions}

We have continued the efficiency calibration of a germanium detector for high-precision $\gamma$-ray spectroscopy. This detector is
used primarily for branching-ratio measurements of super-allowed 0$^+ \rightarrow $ 0$^+$ decays where precisions of the branching ratios
of the order of 0.2\% or better are needed. With the present study, we achieved a precision of the absolute detection efficiency 
for energies down to 40~keV of 0.2\% and of the relative detection efficiency of 0.15\%.

During recent years, we had access to short-lived sources produced at ISOLDE, which we used to add in particular measurements
below 100~keV. We thus added $^{169}$Yb as a new source, which has well-known $\gamma$ rays between 50~keV and 300~keV. In addition, we 
improved previous measurements by increasing the detection statistics or by increasing the statistics of our MC simulations.

Finally, we added a high-energy source, $^{38}$K. This isotope has a single high-energy $\gamma$ ray which allowed us
to verify the total-to-peak ratio above 2~MeV. This source also yielded an additional data point for the single- and double-escape probability
for annihilation radiation.

\section*{Acknowledgment}

We thank K. Johnston and U. K\"oster for their help during the sample collection at ISOLDE.


\end{document}